\documentclass[10pt,twocolumn,letterpaper]{article}

\usepackage{wacv}
\usepackage{times}
\usepackage{epsfig}
\usepackage{graphicx}
\usepackage{amsmath}
\usepackage{amssymb}

\usepackage[caption=false]{subfig}
\usepackage[linesnumbered,ruled,vlined]{algorithm2e}
\usepackage{multirow}
\usepackage{makecell}
\usepackage{comment}
\usepackage{colortbl}

\def\wacvPaperID{10} 

\wacvfinalcopy 

\ifwacvfinal
\def\assignedStartPage{9876} 
\fi

\ifwacvfinal
\usepackage[breaklinks=true,bookmarks=false]{hyperref}
\else
\usepackage[pagebackref=true,breaklinks=true,colorlinks,bookmarks=false]{hyperref}
\fi

\ifwacvfinal
\else
\fi

\begin{document}

\title{Interpretable security analysis of cancellable biometrics using constrained-optimized similarity-based attack}

\author{Hanrui Wang, Xingbo Dong, Zhe Jin\\
Monash University Malaysia\\
47500 Subang Jaya, Selangor, Malaysia\\
{\tt\small \{hanrui.wang, xingbo.dong, jin.zhe\}@monash.edu}
\and
Andrew Beng Jin Teoh\\
Yonsei University, South Korea\\
50 Yonsei-ro Seodaemun-gu, Seoul, South Korea\\
{\tt\small bjteoh@yonsei.ac.kr}

\and
Massimo Tistarelli\\
University of Sassari, Italy\\
Alghero, SS 07041, Italy\\
{\tt\small tista@uniss.it}
}

\maketitle
\pagestyle{empty}
\thispagestyle{empty}

\begin{abstract}
In cancellable biometrics (CB) schemes, template security is achieved by applying, mainly non-linear, transformations to the biometric template. The transformation is designed to preserve the template distance/similarity in the transformed domain.
Despite its effectiveness, the security issues attributed to similarity preservation property of CB are underestimated. Dong et al. [BTAS'19], exploited the similarity preservation trait of CB and proposed a similarity-based attack with high successful attack rate. The similarity-based attack utilizes preimage that are generated from the protected biometric template for impersonation and perform cross matching. In this paper, we propose a constrained optimization similarity-based attack (CSA), which is improved upon Dong's genetic algorithm enabled similarity-based attack (GASA). The CSA applies algorithm-specific equality or inequality relations as constraints, to optimize preimage generation. We interpret the effectiveness of CSA from the supervised learning perspective. We identify such constraints then conduct extensive experiments to demonstrate CSA against CB with LFW face dataset. The results suggest that CSA is effective to breach IoM hashing and BioHashing security, and outperforms GASA significantly. Inferring from the above results, we further remark that, other than IoM and BioHashing, CSA is critical to other CB schemes as far as the constraints can be formulated. Furthermore, we reveal the correlation of hash code size and the attack performance of CSA.
\end{abstract}

\section{Introduction}
\par In the past decade, biometric technology has been widely deployed for identity management systems due to its usability. As a consequence, the proliferation of centralized biometric databases is unavoidable, thus increasing security and privacy concerns.
Cancelable biometrics (CB) is devised to protect the biometric template. CB employs a parameterized, renewable, and irreversible but similarity preserving transformation to convert a biometric template into a protected instance. Due to the property of similarity preservation, templates comparison can be done in the transformed domain while prohibiting the comparison of templates between original and its transformed version \cite{patel2015cancelable}. Since CB was introduced, a large number of methods have been reported. Some representative instances are BioHashing \cite{teoh2006random}, bloom filter \cite{rathgeb2013alignment}, and Index-of-Max (IoM) hashing \cite{jin2017ranking}. Ideally, the following four requirements should be satisfied by any CB scheme:
\begin{itemize}
\item \textit{Non-invertibility or Irreversibility:} It should be computationally hard to retrieve original biometric features from a single or multiple compromised templates, even under the case where the transformation parameters e.g. token are known. Hence, compromise of protected templates will not lead to the privacy invasion.
\item \textit{Revocability or Renewability:} It should be feasible to re-issue a new protected template when the old one is compromised. Other protected templates should remain unchanged when the replacement is carried out.
\item \textit{Non-linkability or Unlinkability:} It is impossible to perform cross matching among multiple protected templates across various applications. When one or multiple protected biometric templates have been stolen, the templates that derived from the same identity that stored in other databases should not be traced.
\item \textit{Performance preservation:} The accuracy performance should not be degraded significantly after transformation. To meet this criterion, the CB scheme has to satisfy the similarity preservation property where the pair-wise distance of templates before and after transformation should be largely preserved. 
\end{itemize}


\par Despite CB has been proved useful, it is vulnerable to various attacks. Firstly, CB could be attacked by attack via record multiplicity (ARM) (also known correlation attack) \cite{kaur2018random}, which leads to the violation of the unlinkability criteria. The ARM refers to the case when multiple protected templates are stolen, then attackers may correlate those templates that are generated from the same user for cross matching and/or for template generation. Secondly, CB could be harmed by dictionary based attack \cite{wu2018cancelable} and brute force \cite{bolle2002biometric} attack, which result in violating the irreversibility criteria. Our focus in this paper is similarity attack (SA) or also known as pre-image attack \cite{lee2009inverse}.  Similarity attack exploits the similarity preserving property of CB to generate a pre-image from the protected template for impersonation as well as cross-match with the templates in various applications. The preimage is an instance despite not close to the original biometric template, after the transformation, it is highly alike to the targeted protected template. Therefore, SA is harmful to unlinkability.

\par SA attack has been explored earlier in the literature. Most of the work focus on SA against BioHashing assuming its parameter (secret token) is known \cite{patel2015cancelable}. Despite a few proposals such as simple data dimension reduction and permutation, attempted to rectify this issue \cite{lumini2007improved}, vulnerability to SA remains critical \cite{lee2009inverse,dong2019genetic}. Most of the existing SA schemes, such as Genetic algorithm enabled similarity-based attack (GASA) solely exploit similarity preserving property of CB to launch the attack \cite{dong2019genetic,lacharme2013preimage}. However, for some CB schemes, besides distance similarity, some other information can be learned from the protected templates. For instance, in IoM hashing, a set of inequalities can be established from each individual hash code \cite{jin2017ranking,dong2018generalized}. We prove in section \ref{exp_result} that when these inequalities are utilized as constraints to compute a preimage, the discrepancy between the two protected templates that transformed from the preimage and from the original biometric template can be eliminated.

\par In this paper, we propose an attack method called constrained-optimized similarity-based attack (CSA). CSA is built upon the GASA  \cite{dong2019genetic} with augmented algorithm-specific equality or inequality relations as constraints. We explain the attempt to seek a preimage generation problem as a form of supervised learning problem (section \ref{fitnessfunction}) where the constraints are leveraged to guarantee a minimal loss value (eq. \ref{eq:fr} in section \ref{fitnessfunction}), e.g. minimize to zero for IoM hashing. Therefore, an optimal preimage can be generated. With better preimage quality, the successful attack rate can be elevated remarkably. Since preimage generation is based on the supervised learning approach, overfitting is inevitable. We demonstrate that by enlarging the hash code size, the overfitting problem can be alleviated. With large hash code size, the effectiveness of CSA over GASA is observed. We conduct the experiments to assess the security of IoM hashing under LFW dataset via CSA. Likewise, we conduct extensive evaluation on BioHashing to prove the CSA is generalised for other CB. We share our code at https://github.com/azrealwang/csa.

\par The main contributions of this paper are:
\begin{itemize}
\item We propose a constrained optimization based similarity attack abbreviated as CSA. The CSA can be effective to any CB schemes where equality or inequality relations can be formulated.
\item We interpret preimage generation problem from the supervised learning perspective, which justifies why CSA outperforms GASA significantly.
\item We conduct extensive experiments to demonstrate CSA against IoM hashing and BioHashing with LFW face dataset. We reveal the hash code size is one of the key factors to impact attack performance. We also discuss the implication of CSA on security, parameters setting and efficiency with respect to the attack performance.

\end{itemize}

\section{Related work}
\label{relatedwork}

\subsection{Similarity-based attack (SA)}
\label{review_SA}

\par There are several  SA schemes proposed in the literature. Dong et al. \cite{dong2019genetic} and Lacharme et al. \cite{lacharme2013preimage} propose a SA scheme using genetic algorithm (GA). The GA-enabled SA generates preimage with an iterative procedure of selection, mutation and crossover. The experimental results attest that GA-enabled SA is effective to breach Biohashing based cancellable face and fingerprint templates.

Feng et al. \cite{feng2014masquerade} propose a SA scheme based on neural networks. Feng et al. claim that there are several limitations in many SA schemes such that: 1) They rely on less-realistic assumptions, 2) They rely on knowledge of e.g. transformation algorithm and its parameters. Hence, Feng et al. consider two distinct scenarios depending on the availability of transformation algorithm knowledge. With that knowledge, the attack can be launched for preimage generation. If transformation algorithm knowledge is not available, the attack also can be launched by using a multilayer perceptron network and a customized hill climbing \cite{gomez2012face,adler2003sample}.

Kaplan et al. \cite{kaplan2017known} propose a SA scheme by utilizing the relation of multiple compromised transformed templates from the same individual. The scheme exploits distance preserving trait of the transformation algorithm. However, Kaplan's scheme does not apply to the systems that only store one single protected template of each user .

Ghammam et al. \cite{ghammam2020cryptanalysis} propose a constrained optimization SA against IoM hashing. The method applies inequalities as constraints to seek preimage solution. 

\subsection{IoM hashing}
\label{review_iom}
\par The IoM hashing is a ranking-based locality sensitive hashing method designed for biometric template protection. The IoM hashing transforms a biometric template into a set of top ranked discrete indices of random subspace projection as hash code \cite{jin2017ranking}. Firstly, biometric feature \(\textbf{x} \in \mathbb{R}^N\) is extracted from the biometric data. A set of random matrices \(\textbf{R}_l \in \mathbb{R}^{K \times N}, l=1,...,L\), where $L$ is the hash code size and $K$ is the subspace dimension, are drawn from a standard Gaussian distribution \(\mathcal{N}(0,1)\). Hence, each
random projected vector \(\textbf{z}_l \in \mathbb{R}^K\) can be computed as:
\begin{equation}
\label{eq:IoM}
\textbf{z}_l=\textbf{R}_l\textbf{x}.
\end{equation}
For each $\textbf{z}_l$, the hash code \(h_l \in [1,K]\)  can be obtained from the first ranked index of vector $\textbf{z}_l$. The IoM hash vector is then formed as \(\textbf{h}=\{h_l|l=1,...,L\}\). At verification stage, the hamming distance is calculated between $\textbf{h}(\textbf{x})$ and $\textbf{h}(\textbf{y})$ \cite{dong2020open} and decision can be made by comparing the hamming score with respect to a pre-determined threshold value.

\subsection{BioHashing}
BioHashing relies on biometric feature \(\textbf{x} \in \mathbb{R}^N\) extracted from the biometric data and a user-specific tokenized random vector \(\textbf{b}_l \in \mathbb{R}^N, l=1,...,L (L \leq N)\), where $L$ is hash code size \cite{teoh2006random}. A binary discretization is applied to compute a BioHash code $h_l$ as 
\begin{equation}
h_l=Sig(\textbf{x}^T\textbf{b}_l-\tau),
\label{eq:biohash}
\end{equation}
where \(Sig(\cdot)\) is defined as a signum function and $\tau$ is an empirically determined threshold. 

\section{Constrained-optimized similarity-based attack (CSA)}
\label{proposal}
In this section, the proposed CSA is presented. We begin with a background introduction of similarity-based attack in section \ref{SA}. Then, we address the loss function that we use to optimize the preimage search in section \ref{fitnessfunction}. Then we reveal the preimage generation with the proposed constrained optimization technique can be analyzed from the supervised learning perspective in section \ref{co}. Thereafter, as a toy example, IoM hashing is used to evaluate the CSA, particularly, the inequality constraints of IoM hashing is presented in section \ref{ineq}. Note that in section \ref{proposal} BioHashing algorithm will not be discussed but it will be assessed in experiments to prove the generalization (section \ref{exp_result}). At last, a preimage generation algorithm, namely Augmented Lagrangian Genetic Algorithm (ALGA) is given in section \ref{ALGA}.

\subsection{Similarity-based attack (SA) framework}
\label{SA}
A given CB transformation function $f(\cdot)$ could be defined as 
\begin{equation}
\label{eq:h}
\textbf{y}=f(\textbf{x}).
\end{equation}
where $\textbf{x}$ represents the biometric template and $\textbf{y}$ represents protected template. 

\par Ultimately, the aim of SA is to generate a preimage $\hat{\textbf{x}}$ \cite{chen2019deep} of $\textbf{y}$. Thus, SA is intuitively an optimization task to search optimal $\hat{\textbf{x}}$:
\begin{equation}
\label{eq:SA_x}
\arg \min_{\hat{\textbf{x}}} \|\textbf{x}-\hat{\textbf{x}}\|.
\end{equation}
In a CB system, $\textbf{x}$ is discarded after transformation, however, the transformed templates $\textbf{y}$ are stored. In order to preserve accuracy, the relative distances between $\textbf{y}$ and \(\hat{\textbf{y}}=f(\hat{\textbf{x}})\) are to be preserved. Thus, e.q. \ref{eq:SA_x} can be reformulated to 
\begin{equation}
\label{eq:SA_h}
\arg \min_{\hat{\textbf{x}}} D(\textbf{y},\hat{\textbf{y}}),
\end{equation}
where $D(\cdot)$ is an algorithm-specific distance function.

\subsection{Loss function}
\label{fitnessfunction}
The preimage generation that required by CSA can be perceived as a supervised learning problem, where the goal is to learn optimal preimage by utilizing labeled data as a supervising signal. Note that in CSA context, learning/training refers to the preimage generation but not to model training under conventional supervised learning setting. In supervised learning, a training data set {$\textbf{X}$, $\textbf{Y}$ }  is to be availed where $\textbf{X}$  is input data and  $\textbf{Y}$  is the label of $\textbf{X}$. In our context, $f(\textbf{X})$ in eq. (2) can be seen as a predicted label. Based on eq. \ref{eq:h} and eq. \ref{eq:SA_h}, the loss function of CSA preimage generation $\mathcal{L}(\cdot)$ is defined as
\begin{equation}
\label{eq:f}
\mathcal{L}=D(\textbf{Y},f(\textbf{X})).
\end{equation}
where $D(\cdot)$ is an algorithm-specific distance function.

Specifically for IoM hashing as an example, biometric template $\textbf{x}$ is \textit{unknown} as it is discarded after transformation, while user-specific matrix $\textbf{R}^\ast=\{\textbf{R}^\ast_l\}_{l=1}^L$ and the protected template $\textbf{h}=\{h_l\}_{l=1}^L$ are \textit{available} as they are stored in the database. Recall $h_l$ is the index value of maximum entry of $\textbf{R}^\ast_l\textbf{x}$ (IoM hash code) and $L$ is the hash code size of IoM hashing. The optimal preimage $\hat{\textbf{x}}$ is the target of searching. Furthermore, for each $\textbf{R}_l^\ast\textbf{x}$, there is a corresponding "label", i.e. $h_l$  from supervised learning perspective. As $\textbf{R}_l^\ast$ is \textit{known} and $\textbf{x}$ is \textit{unknown}, the {$\textbf{X}$, $\textbf{Y}$ }  in eq. \ref{eq:f} can be defined as 
\begin{equation}
\label{eq:f_para}
\begin{aligned}
    &\textbf{X}=\textbf{R}^\ast=\{\textbf{R}^\ast_l\}_{l=1}^L;\\
    &\textbf{Y}=\textbf{h}=\{h_l\}_{l=1}^L.\\
\end{aligned}
\end{equation}
Therefore, there are $L$  pairs of $\{(\textbf{R}^\ast_l,h_l)\}_{l=1}^L$ can be formed.

Finally, the loss function in eq. \ref{eq:f} can be reformulated as:
\begin{equation}
\label{eq:fr}
\mathcal{L}(\hat{\textbf{x}})=D(\textbf{h},f(\textbf{R}^\ast)),
\end{equation}

\subsection{Constrained optimization}
\label{co}
In this subsection, the reasons behind the higher attack success rate of the CSA model are discussed and explained. Moreover, the correlation between the hash code size and the security of CB, already addressed in [10], is further analyzed under the perspective of supervised learning. As it will be demonstrated, a larger hash code size leads to lower security.

Let \(\textbf{h}=f(\textbf{x};\textbf{R}^\ast)\) and \(\hat{\textbf{h}}=f(\hat{\textbf{x}};\textbf{R}^\ast)\) be the IoM hash vector that generated from biometric template ${\textbf{x}}$ and its counterpart that generated from pre-image $\hat{\textbf{x}}$ during "training" stage. 
Then suppose \(\textbf{h}^\prime=f(\textbf{x};\textbf{R}^{\ast^\prime})\) and \(\hat{\textbf{h}}^\prime=f(\hat{\textbf{x}};\textbf{R}^{\ast^\prime})\)  are the IoM hash vectors for verification where $\textbf{R}^{\ast\prime}$  is a new user-specific random matrix for $\textbf{h}^\prime$ and $\hat{\textbf{h}}^\prime$  {($\textbf{R}^{\ast\prime} \neq \textbf{R}^{\ast}$)}. For the sake of clarity, we define several terms:

\begin{itemize}
\item Definition 1: \textit{Test error (TEE)}, or prediction error, refers to the deviation of true $\textbf{h}^\prime$ and predicted $\hat{\textbf{h}}^\prime$ from testing data. It 
is used to estimate the attack performance of unseen data. As they are integer vectors, TEE can be computed with
\begin{equation}
    \label{eq:tee}
    TEE=\frac{1}{n}\sum^n_{i=1}[D(\textbf{h}^\prime_i, \hat{\textbf{h}}^\prime_i)]^2.
\end{equation}
where $D(\cdot)$ here is the hamming distance and $n$ is the repetitive verification times. In our experiment, we set $n=10$  (section \ref{protocol}). $\textbf{h}^\prime_i$ and $\hat{\textbf{h}}^\prime_i$ are generated by different $\textbf{R}^{\ast\prime}$ for each verification.
\item Definition 2: \textit{Training error (TRE)} refers to the deviation of true $\textbf{h}$ and predicted $\hat{\textbf{h}}$  \cite{alpaydin2014introduction}. Similarly, TRE can be computed with:
\begin{equation}
    \label{eq:tre}
    TRE=[D(\textbf{h},\hat{\textbf{h}})]^2.
\end{equation}
\item Definition 3: \textit{Variance (VAR)} indicates the mean square difference of  $\hat{\textbf{h}}^\prime$ and the expected value of $\hat{\textbf{h}}$ \cite{alpaydin2014introduction}. Under supervised learning setting, VAR can be regarded as a measure of system overfitting. The larger VAR, the more system prone to overfitting.  However, since these two values are computed with distinct user-specific matrices, so they are incomparable. Thus, VAR is approximately defined as
\begin{equation}
    \label{eq:var}
    VAR \approx \frac{1}{n}\sum^n_{i=1}[D(\textbf{h}^\prime_i, \hat{\textbf{h}}^\prime_i)-D(\textbf{h},\hat{\textbf{h}})]^2.
\end{equation}
\end{itemize}

To gain a decent attack performance (indicated by TEE), both small TRE and VAR are expected \cite{alpaydin2014introduction}. 

\begin{figure}[!t]
\centering
\includegraphics[width=3.2in]{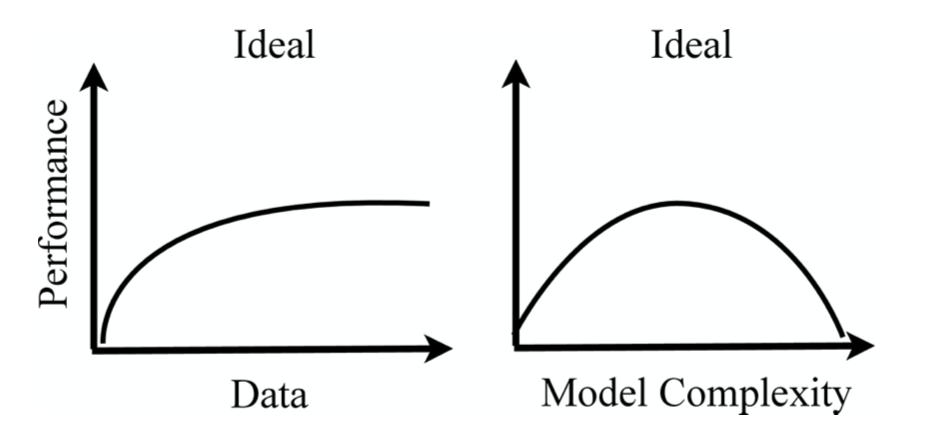}
\caption{Performance related to dataset size and model complexity under ideal conditions \cite{zhu2016we}.}
\label{fig:ideal}
\end{figure}

In general, overfitting is determined by the two factors, namely training data size and model complexity. As shown in Fig. \ref{fig:ideal}, in an ideal setting, the test accuracy increases with respect to data size due to good generalization (no overfitting) whereas the test accuracy will deteriorate when the model complexity is overly high due to overfitting. In CSA context, the model complexity is fixed as CSA model does not have learnable parameters. The overfitting issue of the CSA relies solely on IoM hash code size , $L$ , which is equivalent to training data size. Thus, the solution to alleviate the overfitting in the CSA is just to increase the $L$ .

Based on eq. \ref{eq:tre}, TRE is smaller when $\textbf{h}$ and $\hat{\textbf{h}}$ are closer. Ideally, $TRE=0$ when $\hat{\textbf{h}}=\textbf{h}$. The constrained optimization model that we designed for CSA is based on the algorithm-specific inequality or equality relations. The constraints are established to force $\hat{\textbf{h}}$  be closer to or be (ideally) identical to $\textbf{h}$. The details of the algorithm-specific constraints designed for IoM hashing are presented in section \ref{ineq}.

\subsection{Inequality constraints of IoM hashing}
\label{ineq}
The constraints are established to minimize eq. \ref{eq:fr}, ideally $\mathcal{L}(\cdot)=0$. Equation \ref{eq:fr} can still converge without the constraints but it would not be optimal.

In IoM hashing, each hash code is derived from the first ranked index after projecting the biometric template onto a random subspace \cite{jin2017ranking}. This facilitates the construction of the inequality relations, which can be exploited as the constraints for CSA. If the constraints are satisfied, the TRE is equal to zero as \(\hat{\textbf{h}} =\textbf{h}\). 

\par Let   \(\hat{\textbf{z}}=[\hat{z}_{1},\hat{z}_{2},...,\hat{z}_{K}]=\textbf{R}^\ast\hat{\textbf{x}}\)  be the random projected vector of preimage and $\hat{h}$  be the first ranked index value of $\hat{\textbf{z}}$. In the event of CSA, user-specific matrix $\textbf{R}^\ast$ and $h$ are compromised. The attacker could leverage the clue $\hat{h}=h$ to minimize the hamming distance. Based on the knowledge of $\textbf{R}^\ast$ and $h$ , the inequality relations could be established as
\begin{equation}
\label{eq:inequality}
\hat{z}_{h} \geq \forall \hat{z}_{k} \in \hat{\textbf{z}}, k=1,...,K.
\end{equation}
A constraint function $c(\cdot)$ can be defined based on eq. \ref{eq:inequality} as 
\begin{equation}
\label{eq:ineqf}
c(\hat{\textbf{x}})=\{\hat{z}_{k}-\hat{z}_{h} |k=1,..,K\}, where~\hat{z}_{k} \in \hat{\textbf{z}}=\textbf{R}^\ast\hat{\textbf{x}}.
\end{equation}
Thus, the inequality constraint is \(c(\hat{\textbf{x}}) \leq 0\). If the constraint is unsatisfied, we name it as a violation. To quantify the violation, a constraint violation function $v(\cdot)$ is devised as follow:
\begin{equation}
\label{eq:violation}
v=\max c(\hat{\textbf{x}}).
\end{equation}
In general, the constraint is satisfied if $v(\cdot) \leq \tau$, where $\tau \approx 0$, or \(\tau=0\) ideally. When $v(\cdot) \leq 0$ is satisfied (our case), $\mathcal{L}(\hat{\textbf{x}})=0$.

\subsection{Augmented Lagrangian Genetic Algorithm}
\label{ALGA}
The original SA is an unconstrained optimization problem based on eq. \ref{eq:fr} to approximate $\hat{\textbf{x}}$, which is kind of ad-hoc. GA is employed in GASA as a searching algorithm \cite{dong2018generalized}. In the traditional GA, the output is produced with a repeated procedure of selection, mutation and crossover until convergence (i.e. meeting stopping criteria). However, traditional GA is limited to solve unconstrained optimization problems \cite{adeli1994augmented}. In this paper. we adopt ALGA \cite{conn1997globally}  as a means of searching algorithm to solve $\hat{\textbf{x}}$ based on the constraints formulated in eq. \ref{eq:fr} and eq. \ref{eq:ineqf}. ALGA is an iterative 4-steps procedure (one generation) as follows:
\begin{itemize}
\item \textit{Step 1:} Combine the loss function $\mathcal{L}(\cdot)$ (eq. \ref{eq:fr}) and constraint function $c(\cdot)$ (eq. \ref{eq:ineqf}) in terms of Lagrangian barrier function $LB(\cdot)$. The goal is to transform a constrained to unconstrained problem, which is applicable to GA. For the definition of Lagrangian barrier function and detailed combination algorithm please refer to \cite{conn1997globally}.
\item \textit{Step 2:} Minimize $LB(\cdot)$ with GA. 
\item \textit{Step 3:} The optimization process will be terminated when \(|\mathcal{L}_{n-1}(\cdot)-\mathcal{L}_n(\cdot)| \leq \tau_1\)  and \(v(\cdot) \leq \tau_2\) where $n$ is the generation number, $\tau_1$ and $\tau_2$ are the selected tolerance where $\tau_\ast \approx 0$, or \(\tau_\ast=0\),  *={1, 2} ideally. 
\item \textit{Step 4:} Go to \textit{Step 1} and recreate $LB(\cdot)$ referring to \cite{conn1997globally}, if stopping criteria is not satisfied. It indicates that the inequality constraint must be satisfied when ALGA converges, while the constraint might not be satisfied at every generation.
\end{itemize}

 A complete ALGA algorithm description can be found at algorithm \ref{alg:ALGA} \cite{conn1997globally}.
 
Other than ALGA, there are more optimization methods where the constraints can be applied. In this work, we are more interested in the constrained optimization than the searching algorithm, so ALGA is only the choice of implementation.


\begin{algorithm}[!t]
\DontPrintSemicolon
\SetKwInOut{Input}{Input}\SetKwInOut{Output}{Output}
 \caption{Augmented Lagrangian Genetic Algorithm}
 \label{alg:ALGA}
 \Input{Compromised template $\textbf{h}$ and user-specific $\textbf{R}^\ast$, loss function $\mathcal{L}(\cdot)$, constraint function $c(\cdot)$, constraint violation function $v(\cdot)$, stopping criteria $\tau_1$, $\tau_2$}
 \Output{Preimage $\hat{\textbf{x}}$}
 \BlankLine
 Initialization: Initial $\hat{\textbf{x}}$ with random standard Gaussian vector\;
 \Repeat{\(|\mathcal{L}_{n-1}(\cdot)-\mathcal{L}_n(\cdot)| \leq \tau_1\) and \(v(\cdot) \leq \tau_2\)}{
    Complete Lagrangian barrier function $LB(\cdot)$ by combining $\mathcal{L}(\cdot)$ and $c(\cdot)$ refer to \cite{conn1997globally}\;
    Minimize $LB(\cdot)$ with GA\;
    Estimate if stopping criteria is met\;
    }
\end{algorithm}

\section{Experiments}

\subsection{Dataset}
LFW dataset \cite{huang:inria-00321923} is used to evaluate the CSA on IoM hashing. There are 5749 users in the entire LFW dataset. The subjects with more than 10 images are chosen and the first 10 images of each subject are selected for our experiments. The subset of LFW dataset consists of 1580 images which contains 158 users and 10 images for each user. The face feature is extracted into a 512-dimensional vector by a deep learning model, InsightFace \cite{deng2019arcface}.


\begin{table*}[!t]
\renewcommand{\arraystretch}{1.3}
\caption{CSA performance with subspace dimension \(K=16\).}
\label{tab:L}
\centering
\setlength{\tabcolsep}{2.7mm}{
\begin{tabular}{|c|c|c|c|c|c|c|c|c|c|}
\hline \multirow{2}{*}{\begin{tabular} [c]{@{}l@{}}Hash code\\ size ($L$)\end{tabular}}&\multicolumn{3}{c|}{IoM}&\multicolumn{3}{c|}{GASA}&\multicolumn{3}{c|}{Proposed CSA}\\
\Xcline{2-10}{0.4pt} &$\theta^2$&EER(\%)&FMR(\%)&SAR(\%)&FAI(\%)&TEE&SAR(\%)&FAI(\%)&TEE\\
\hline
\hline 8&0.7639&14.40&9.78&10.94&1.16&0.8537&11.06&1.28&0.8491\\
\hline 16&0.7639&8.59&9.44&12.41&2.97&0.8373&11.14&1.70&0.8491\\
\hline 32&0.7101&3.87&2.69&3.96&1.27&0.8338&4.83&2.14&0.8311\\
\hline 64&0.7351&2.00&2.24&5.40&3.16&0.8139&12.03&9.79&0.7740\\
\hline 128&0.7379&1.10&1.10&7.54&6.44&0.7823&42.13&41.03&0.6693\\
\hline 256&0.7455&0.72&0.71&24.11&23.39&0.7263&92.08&91.36&0.5023\\
\hline 512&0.7525&0.60&0.61&64.12&63.51&0.6482&\textbf{99.19}&\textbf{98.58}&\textbf{0.2866}\\
\hline 
\end{tabular}
}
\end{table*}

\begin{figure*}[!t]
\centering
\subfloat[GASA]{
\includegraphics[width=3.2in]{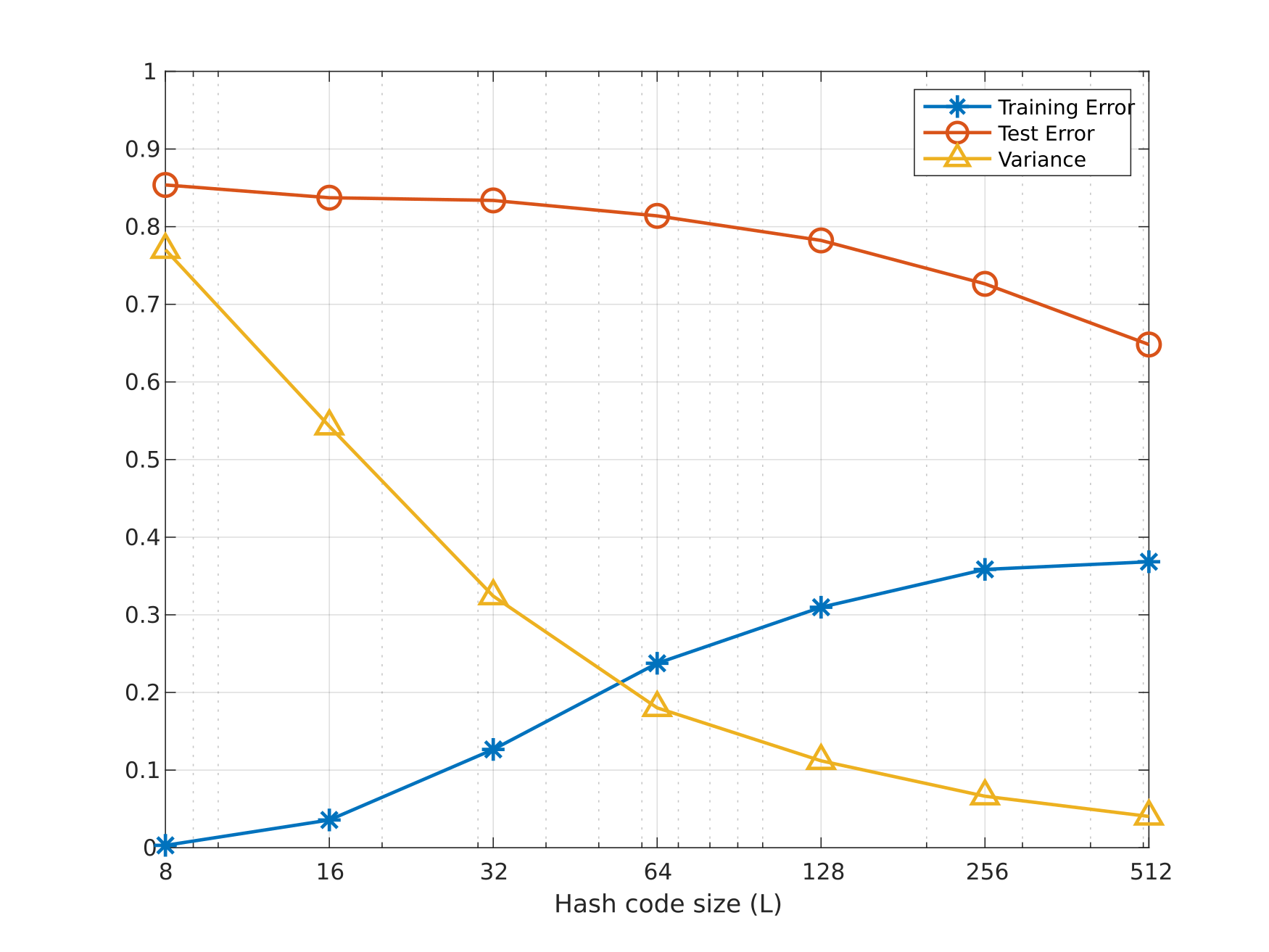}
\label{fig:SA_err}
}
\subfloat[CSA]{
\includegraphics[width=3.2in]{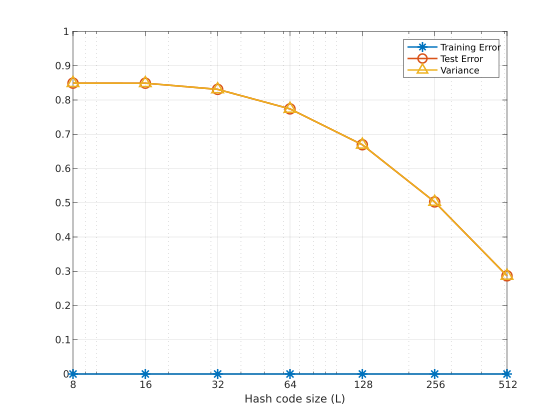}
\label{fig:CSA_err}
}
\caption{Error of GASA and CSA with different hash code size $L$.}
\label{fig:err}
\end{figure*}

\begin{table}[!t]
\renewcommand{\arraystretch}{1.3}
\caption{CSA efficiency with subspace dimension \(K=16\).}
\label{tab:efficiency}
\centering
\setlength{\tabcolsep}{0.5mm}{
\begin{tabular}{|c|c|c|c|c|}
\hline \multirow{2}{*}{\begin{tabular} [c]{@{}l@{}}Hash code\\ size ($L$)\end{tabular}}&\multicolumn{2}{c|}{GASA}&\multicolumn{2}{c|}{Proposed CSA}\\
\Xcline{2-5}{0.4pt} &Generations&Time (s)&Generations&Time (s)\\
\hline
\hline 16&68&11.88&5&142.95\\
\hline 32&87&15.78&5&182.00\\
\hline 64&109&26.13&5&201.87\\
\hline 128&111&34.10&5&364.96\\
\hline 512&94&97.84&5&3661.63\\
\hline 
\end{tabular}
}
\end{table}

\begin{table*}[!t]
\renewcommand{\arraystretch}{1.3}
\caption{CSA performance against Biohashing.}
\label{tab:biohashing}
\centering
\setlength{\tabcolsep}{3.7mm}{
\begin{tabular}{|c|c|c|c|c|c|c|}
\hline \multirow{2}{*}{\begin{tabular} [c]{@{}l@{}}Hash code\\ size ($L$)\end{tabular}}&\multicolumn{2}{c|}{Biohashing}&\multicolumn{2}{c|}{GASA}&\multicolumn{2}{c|}{Proposed CSA}\\
\Xcline{2-7}{0.4pt} &EER(\%)&FMR(\%)&SAR(\%)&FAI(\%)&SAR(\%)&FAI(\%)\\
\hline
\hline 512&0.63&0.63&34.43&33.80&\textbf{72.74}&\textbf{72.11}\\
\hline 
\end{tabular}
}
\end{table*}

\subsection{Experiment protocol}
\label{protocol}
In our experiments, we assume only one protected template is stored in the database for each subject, and this template is transformed from the first sample of each subject in the dataset. The remaining nine samples of each subject are used for verification. This protocol is to evaluate whether the generated preimage is effective to breach the system after the user re-issues a protected template using new face image. Furthermore, we adopt distinctive $\textbf{R}^\ast$ in our experiments for preimage generation and verification. This aims to simulate re-issue of protected template scenario and evaluate the possibility of preimage to breach multiple systems. The hash code size $L$, which varies from 8 to 512 for IoM hashing is examined and the preimage generation process under each scenario of selected $L$ is executed only once due to computational cost. 

\par For attack performance assessment, we propose two new metrics, namely successful attack rate (SAR) and false acceptance increment (FAI). \textit{SAR} are recorded using acceptance rate of preimage $\hat{\textbf{x}}$. By excluding false match rate (FMR), \textit{FAI}=SAR-FMR, indicates true attack performance. A higher FAI implies a more effective attack performance. In addition, equal error rate (EER) is computed to evaluate the performance of IoM hashing with respect to $L$. All performance assessments are repeatably executed ten times with different $\textbf{R}^\ast$.

\par For efficiency evaluation, the number of generations and computation cost in second are recorded when preimage generation of one sample reaches convergence. The machine we used for simulation is equipped with MATLAB Ver. 2019b, 2.7 GHz Dual-Core Intel Core i5 CPU and 1867 MHz 8GB RAM.

\subsection{Experiment results}
\label{exp_result}
We first investigate the effect of $L$  with respect to the attack performance of CSA.  For each $L$, threshold $\theta$ (for the convenience of comparison, it is recorded as $\theta^2$), EER and FMR of IoM hashing, SAR, FAI and TEE of GASA and the proposed CSA are reported in Table \ref{tab:L}. The results show that when $L \leq 32$, TEE of CSA is more than 0.83 and it is considerably higher than $\theta^2$ (0.83 vs 0.71). This implies CSA is not so effective against IoM hashing when hash code size is small. However, this is not practical because small hash code size is vulnerable to other security attacks such as brute force attack. On the other hand, when $L>64$, TEE of CSA is perceivably reduced and FAI raises dramatically. When $L=512$, SAR of CSA achieved up to 99.19\% and FAI of CSA achieved 98.58\%. To benchmark GASA, the attack performance of CSA is close to GASA when $L$ is small (e.g. $L \leq 32$). However, when $L$ increases, TEE of CSA reduces much faster than TEE of GASA. Along with SAR and FAI, we can observe that CSA largely outperforms GASA in terms of SAR (CSA: 99.19\% vs GASA: 64.12\%) and FAI (CSA: 98.58\% vs GASA: 63.51\%).

\par The experiment results obtained above are indeed consistent with the analysis in section \ref{proposal}. Figure \ref{fig:err} illustrates the error trend of GASA and CSA. Specifically, when $L$ is small, it results high VAR due to overfitting (yellow curve) (section \ref{co}). Despite TRE (blue curve) of both GASA and CSA are low under this case, TEE (red curve) is still high. This leads to poor attack performance. On the other hand, when $L$ increases, VAR for both GASA and CSA is reduced. The TRE of CSA remains at zero attributed to the inequality constraints, yet TRE of GASA increases due to the increment of $L$. When $L$ is large enough (e.g. $L=512$), the attack performance is mainly affected by TRE. With additional constraints, TRE in CSA can be zero while this is not the case for GASA. Therefore, we can point out that CSA outperforms GASA when $L$ is large enough.

\par To evaluate the efficiency of CSA, the number of generations and time for convergence in second for one preimage generation are recorded in Table \ref{tab:efficiency}. The results show that the number of generations to convergence for CSA is significantly lesser than that of GASA while the preimage computational cost of CSA is comparably higher. This is due to natural distinction of the traditional GA and ALGA, where one generation produced by ALGA refers to a solution of each Lagrangian barrier function, and this process likely consists of multiple generations produced by the traditional GA. The number of generations for ALGA is hence less. Referring to section \ref{ALGA}, ALGA consists of multiple operations, e.g. Lagrangian barrier function $LB(\cdot)$ and GA computation. The total operations of ALGA is much more than just one single GA, hence more computational cost is somehow expected. This demonstrates a phenomenon where high preimage computational cost trades with a good attack performance in CSA. Nevertheless, even approximately one hour (3661.63 seconds) for a successful attack is feasible and worthwhile indeed.

We also conduct CSA on Biohashing by using LFW face dataset. Table \ref{tab:biohashing} reveals that CSA can largely outperform GASA on Biohashing as well in terms of SAR (CSA: 72.74\% vs GASA: 34.43\%) and FAI (CSA: 72.11\% vs GASA: 33.80\%). Therefore, CSA can applied to CB which could apply equalities or inequalities as constraints. As limited by the paper length, we will not address more details about Biohashing.


\section{Conclusion}
In this paper, we propose a novel similarity-based attack on IoM hashing known as CSA. The experiments conducted under LFW face dataset show that CSA is effective in terms of SAR and FAI against IoM hashing and BioHashing. The CSA demonstrates a superior attack performance (SAR and FAI) over GASA due to the employment of constrained optimization. We further analyze that the constraints in CSA can significantly reduce TRE whereas it is impossible for GASA. The theoretical justification on CSA attack performance is consistent with the empirical observations. We further reveal that 1) the attack performance of the CSA is closely related to the IoM hash code size; 2) the CSA trades the preimage computational cost with superior attack performance, nonetheless, the average computational cost for $L=512$ is approximately one hour, which is acceptable and worthwhile. 

However, the CSA model suffers from three limitations. Firstly, the CSA relies on information leakage restrictively. This means it requests the knowledge of the protected template, transformation algorithm and its associated parameters  (eq. \ref{eq:fr}). Secondly, unlike GASA which requires no constraints, CSA expects constraints to be established, which could be difficult for some cancellable biometric schemes such as Bloom filter. However, CSA might outperform GASA if those constraints can be identified. Finally, the generated preimage (i.e. the feature vector) is not available to reconstruct the face image.


In future developments of this research, the generalization of the CSA model against other CB schemes will be addressed. Moreover, countermeasures to enhance the security of CB against attacks including similarity-based attacks, such as adopting non-linear transformation, or integrating permutations before applying the transformation, will be investigated.


\section*{Acknowledgment}
This work has been partially supported by the European Union COST Action CA16101, University of Sassari fondo di Ateneo per la ricerca 2020, Italian Ministry for Research special research project SPADA, Fundamental Research Grant Scheme (FRGS/1/2018/ICT02/MUSM/03/3) and NVIDIA Corporation donation of the Titan Xp GPU.

{\small
\bibliographystyle{ieeetr}
\bibliography{egpaper}
}

\end{document}